# Forecasting the Volatilities of Philippine Stock Exchange Composite Index Using the Generalized Autoregressive Conditional Heteroskedasticity Modeling

**Novy Ann M. Etac** and **Roel F. Ceballos**

Department of Mathematics and Statistics, College of Arts and Sciences
University of Southeastern Philippines, Bo. Obrero, Davao City
Email : novyannetac@yahoo.com and ceballosroel@gmail.com

**ABSTRACT**

*This study was conducted to find an appropriate statistical model to forecast the volatilities of PSEi using the model Generalized Autoregressive Conditional Heteroskedasticity (GARCH). Using the R software, the log returns of PSEi is modeled using various ARIMA models and with the presence of heteroskedasticity, the log returns was modeled using GARCH. Based on the analysis, GARCH models are the most appropriate to use for the log returns of PSEi. Among the selected GARCH models, GARCH (1,2) has the lowest AIC value and also has the highest LL value implying that GARCH (1,2) is the best model for the log returns of PSEi.*

## 1. INTRODUCTION

The Philippine Stock Exchange Composite Index (PSEi) is the country's official stock exchange and is one of the oldest in terms of establishment in Southeast Asia (PSE Academy, 2011). It is highly observed and monitored (Chen & Diaz, 2014) due to its inclusion as one of the top 16 economies of the world by 2050, as predicted by the Hong Kong and Shanghai Banking Corporation (HSBC). The forecast of the Philippine stock exchange returns is currently a primary interest because of its immense possibility to propel upwards. The forecast of the volatilities of PSEi will be utilized to know how budgets will be allocated for a certain period and to know how to plan the expenses to prevent from certain business downfalls. The Generalized Autoregressive Conditional Heteroskedasticity (GARCH) is an approach that can be used to estimate the volatilities of a time series data. Volatilities are not constant. It can either be high or low depending on the period or it has the tendency to exhibit clustering. Safadi (2017) studied the use ARMA-GARCH model to estimate the volatility of daily stock market indices when volatility clustering is observed. The variations in volatility of a time series data can be analyzed using GARCH models for it can accurately describe the phenomenon of volatility clustering.

## 2. PROCEDURE OF THE STUDY

The daily data of the Philippine Stock Exchange Composite Index are taken from the official website of the PSEi. The data is from January 3, 2011 to July 15, 2016 and has a total of 1349 observation.

### 2.1. Model Identification

To identify the appropriate model for the log-returns of PSEi, test for stationarity and test for heteroskedasticity must be carried out. The following tests are used:

1. The Augmented Dickey–Fuller (ADF) test by Dickey and Fuller (1981) was done to test the stationarity of the log returns of PSEi.
2. Before fitting the GARCH model, we fit an ARIMA model to the log returns ($Y_t$). We test the residuals for the presence heteroskedasticity using the Lagrange Multiplier (LM) test. We then fit the GARCH model to the data if heteroskedasticity is present.
3. The Akaike's Information Criterion (AIC) and Log Likelihood (LL) were used to determine which of the candidate models is the best model for the data.

### 2.2. Estimation of Parameters

The GARCH $(r, s)$ model for the natural log returns $Y_t$ is defined by

$$Y_t = \tau_t \varepsilon_t,$$

$$\tau_t^2 = c + \sum_{i=1}^{r} \alpha_i Y_{t-i}^2 + \sum_{j=1}^{s} \beta_j \tau_{t-j}^2,$$

where

    the parameters $c > 0, \alpha_i \geq 0$, and $\beta_j \geq 0$,

    $r$ and $s$ are the degree of the GARCH and ARCH process respectively,

    $Y_t$ is the log return in terms of time t of the PSEi, and

    $\varepsilon_t$ is a white noise.

Using the R software, the estimates for the parameters $(c, \alpha_i, \beta_j)$ will be obtained for the best fit of the GARCH model. The package tseries and the function garch will be used to estimate these parameters. In fitting the model, the idea of parsimony is important in which the model should have as small parameters as possible but still capable of explaining the series. The more parameters estimated the greater the noise that can be introduced into the model. Hence, in the case of GARCH$(p, q)$ model, $p \leq 2$ and $q \leq 2$ is more suitable to lessen the noise.

### 2.3. Diagnostic Checking

After the identification and estimation of the parameters for the fitted model, the following statistical test will be used in the diagnostic checking of the model.

1.) Test for normality. The normality of the residuals will be checked by plotting the histogram of the residual. To further check for its normality the Kolmogorov-Smirnov test will be conducted. The goodness-of-fit test is an important step for testing the selected model. A good fitted model must produce residuals that are approximately uncorrelated in time.

2.) Test for independence. The independence of the residual will be tested using the Ljung-Box test by Ljung and Box (1978). This is used to test the serial correlation of the residuals of the fitted GARCH model. If the null hypothesis is not rejected, it means that the residuals are independent and uncorrelated. Otherwise, the model needs modification since serial correlation is present.

## 3. RESULTS

### 3.1. Stationarity

The time series plot of the recorded daily Philippine Stock Exchange Composite Index ($Y_t$) is shown in Figure 1. It can be observed that in the first months of 2013 there is a huge increase on the stock exchange, but in mid-2013 a sudden decrease was observed that continued for a few months. An increase and decrease of stock exchange is observed for a few months that continued until 2014. A sudden decrease in the exchange was also observed in mid-2015 but eventually never continued since an increase is observed afterward. The highest peak of the series can be seen in 2015 where the index reached to 4206.01.

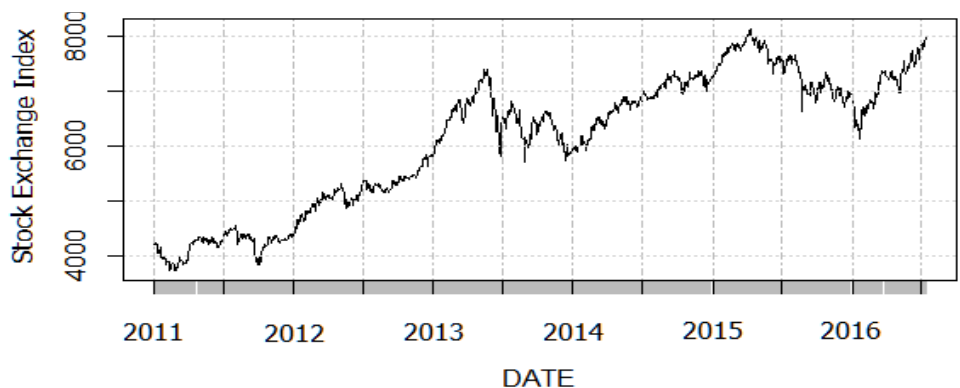

**Figure 1.** Historical Plot of the daily PSEi

It is desirable to have a stationary time series data primarily because the traits of a stationary data set allow one to assume models that are independent of a starting point. However, the plot in Figure 1 shows that the PSEi ($P_t$) is not stationary since the data does not fluctuate around some common mean or location. We confirm this by using the ADF test for stationarity (see Table 1) where the p-value is 0.468 which implies that we cannot reject the null hypothesis (at 5% significance) and hence the data is non-stationary. Instead of analyzing the original data $P_t$ which displays unit-root behavior, we analyze the log-returns ($Y_t$) of the PSEi ($P_t$). The plot of the ($Y_t$) is provided in Figure 2.

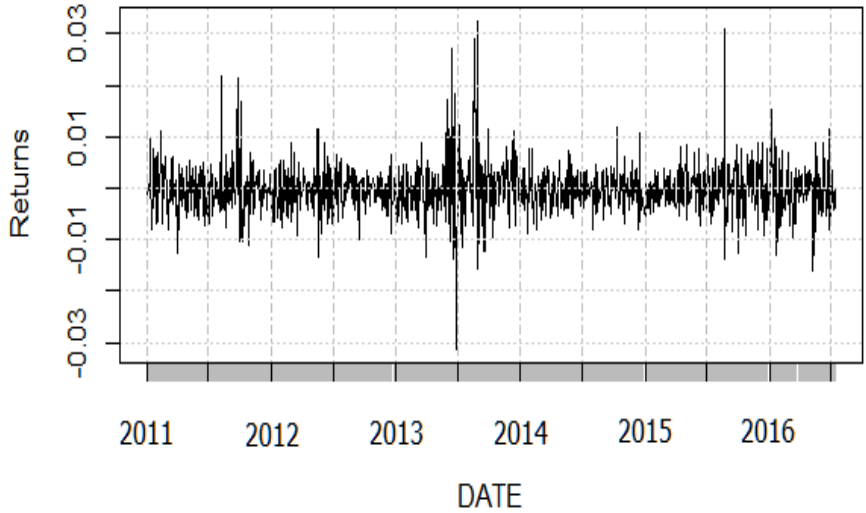

**Figure 2.** Plot of the daily log returns of PSEI

The log-returns $(Y_t)$ is more stationary than $(P_t)$ by merely looking at Figure 2. We confirm this by using the ADF test for stationarity (see Table 1) where the p-value is 0.01 which means reject the null hypothesis (at 5% significance) and hence the data is stationary.

*Table 1.* Augmented Dickey-Fuller Test for Stationarity

| Augmented Dickey- Fuller Test | | |
|---|---|---|
| **Data** | **Value** | **p-value** |
| PSEi ($P_t$) | -2.2593 | 0.468 |
| Log Returns ($Y_t$) | -10.238 | 0.010 |

### 3.2 ARIMA $(p, d, q)$ Models

The correlogram plots of the log returns $(Y_t)$ were presented in Figure 3 and 4. It shows the ACF and PACF plots of the log returns $(Y_t)$. The AFC and PACF can be used as a guide to determine the candidate ARIMA models for the log returns $(Y_t)$.

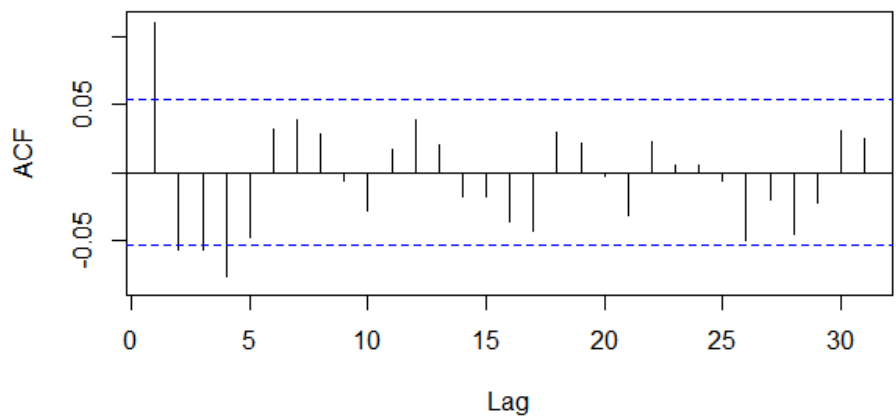

**Figure 3.** ACF Plot of log returns of PSEi

The ACF plot determines the order of $q$ for an ARMA$(p, q)$ model. Observing the plot, the ACF shows a sinusoidal pattern with decreasing intensity and cuts off after lag 1, suggesting an MA (1).

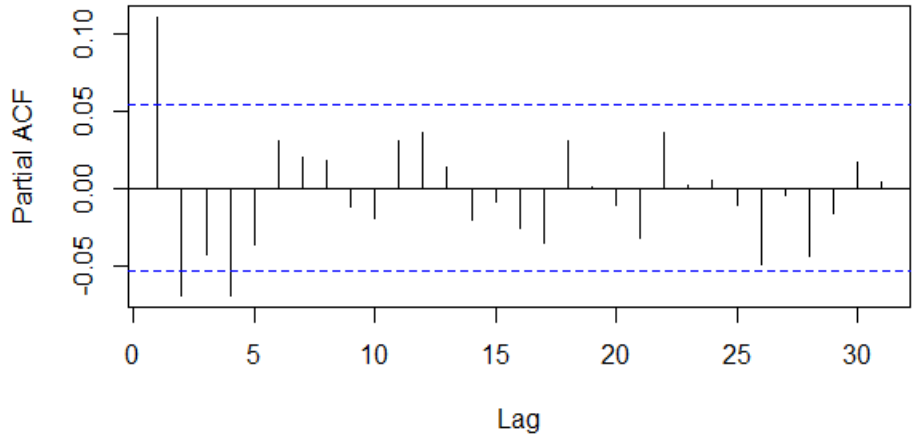

**Figure 4.** PACF Plot of the log returns of PSEI

The PACF plot is used to determine the appropriate order $p$ for an ARMA$(p,q)$ model. The PACF plot cuts off after lag 1 and after the first lag a sinusoidal pattern with decreasing intensity is observed. The AFC and PACF plots suggests an ARMA (1, 1). The parameter estimates of the tentative ARIMA models are presented in Table 2.

*Table 2.* Tentative ARIMA Models of the Log Returns of PSEi

| Models | Constant | AR(1) | MA(1) | MA(2) |
|---|---|---|---|---|
| ARIMA (0,0,1) | $2.083521 \times 10^{-5}$ | - | 0.1236172 | - |
| ARIMA (1,0,0) | $2.086823 \times 10^{-5}$ | 0.1101353 | - | - |
| ARIMA(1,0,1) | $2.080528 \times 10^{-5}$ | -0.2468585 | 0.3665946 | |

### 3.3 Test for Heteroskedasticity

The residuals obtained by fitting each tentative ARIMA models in Table 2 is tested with the presence of heteroskedasticity using the Lagrange Multiplier (LM) test. To qualify the use of GARCH models in the series of the log returns $(Y_t)$, heteroskedaticity must be present in the residuals of the ARIMA models of $(Y_t)$. The LM test value and p-value are provided in Table 3.

*Table 3.* LM Test for the tentative ARIMA models

| Models | LM Test | |
|---|---|---|
| | $x^2$ | p-value |
| ARIMA(0,0,1) | 114.59 | $2.2 \times 10^{-16}$ |
| ARIMA(1,0,0) | 114.36 | $2.2 \times 10^{-16}$ |
| ARIMA(1,0,1) | 116.55 | $2.2 \times 10^{-16}$ |

At a 5% level of significance, the p-value are significant for the LM test of all the tentative ARIMA models. This implies a rejection of the null hypothesis that there is no ARCH effect on the model and proves that there is a presence of heteroskedasticity on the log returns $(Y_t)$ of PSEI.

### 3.4 GARCH$(p, q)$ Models

With the presence of heteroskedasticity observed in the log returns ($Y_t$) of PSEi, the GARCH model is then applied to the series. The estimates for the parameters of these models are then obtained and presented in Table 4 below.

*Table 4.* Estimated Parameters of the Tentative GARCH Models

| GARCH Model | Parameters | | | | |
|---|---|---|---|---|---|
| | $c$ | $\alpha_1$ | $\alpha_2$ | $\beta_1$ | $\beta_2$ |
| GARCH(1,0) | 1.447×10$^{-5}$ | 3.654×10$^{-1}$ | - | - | - |
| GARCH(1,1) | 1.977×10$^{-6}$ | 1.975×10$^{-1}$ | - | 7.151×10$^{-1}$ | - |
| GARCH(1,2) | 2.159×10$^{-6}$ | 2.445×10$^{-1}$ | - | 2.588×10$^{-1}$ | 4.033×10$^{-1}$ |
| GARCH(2,1) | 1.946×10$^{-6}$ | 1.970×10$^{-1}$ | 1×10$^{-8}$ | 7.171×10$^{-1}$ | - |

We fitted a GARCH (1,0), GARCH (1,1), GARCH (1,2), and GARCH (2,1) to the log returns of PSEi and the parameters of these models are presented in Table 4. For GARCH (1,0) model the value of the parameter $c = 0.00001447$, and $\alpha = 0.3654$. For the GARCH (1,1) model the values of the parameter $c = 0.000001977$, $\alpha = 0.1975$, and $\beta = 0.7151$.

### 3.5 Parameter Estimate Analysis

To check if the models are appropriate for the series, Tables 5 to 8 are provided for the analysis of the parameter estimates of the four models.

*Table 5.* Estimate Analysis of GARCH (1,0) Model

| Parameter | Estimate | Standard Error | t value | p-value |
|---|---|---|---|---|
| $c$ | 1.447× 10$^{-5}$ | 7.812× 10$^{-7}$ | 18.522 | 0.00000* |
| $\alpha$ | 3.654× 10$^{-1}$ | 5.492× 10$^{-2}$ | 6.654 | 0.00000* |

*Actual p-value is lower than 0.0001

Table 5 shows that the estimate of parameter $c$ is 0.00001447 with the p-value less than 2×10$^{-16}$, and the estimate of parameter $\alpha$ is 0.3654 with the p-value equal to 2.86×10$^{-11}$. Since both p-values of the two parameters are less than the significant value 0.05, this shows that the GARCH (1,0) model with the specified parameters is a good fit for our data.

*Table 6.* Estimate Analysis of GARCH (1,1) Model

| Parameter | Estimate | Standard Error | t value | p-value |
|---|---|---|---|---|
| $c$ | 1.977× 10$^{-6}$ | 5.089× 10$^{-7}$ | 3.886046 | 0.0000* |
| $\alpha$ | 1.975× 10$^{-1}$ | 3.142× 10$^{-2}$ | 6.286452 | 0.03248 |
| $\beta$ | 7.151× 10$^{-1}$ | 4.501× 10$^{-2}$ | 15.887007 | 0.0000* |

*Actual p-value is lower than 0.0001

The estimate analysis for GARCH (1,1) model is shown in Table 6. This shows that the estimate for parameter $c$ is 0.000001977 with the p-value equal to 0.0000001018, the estimate for parameter $\alpha$ is 0.1975 with the p-value equal to 0.03248, and the estimate for $\beta$ is 0.7151 with the p-value less than 2×10$^{-16}$. All the p-values of the parameters are significant at level 0.05. This shows a strong evidence that the GARCH (1,1) model with the specified parameters is also a good fit for our data.

*Table 7.* Estimate Analysis of GARCH (1,2) Model

| Parameter | Estimate | Standard Error | t value | p-value |
|---|---|---|---|---|
| $c$ | $2.159\times 10^{-6}$ | $5.597\times 10^{-7}$ | 3.858 | 0.00011 |
| $\alpha_1$ | $2.445\times 10^{-1}$ | $3.731\times 10^{-2}$ | 6.553 | 0.0000* |
| $\beta_1$ | $2.588\times 10^{-1}$ | $1.037\times 10^{-1}$ | 2.496 | 0.01254 |
| $\beta_2$ | $4.033\times 10^{-1}$ | $1.085\times 10^{-1}$ | 3.718 | 0.00020 |

*Actual p-value is lower than 0.0001

The estimate analysis for GARCH (1,2) model is shown in Table 7. As shown in the result, all the p-value of the parameters are significant at level 0.05. This shows that the GARCH (1,2) model with the parameters $c, \alpha_1, \beta_1$, and $\beta_2$ is also a good fit for our data.

*Table 8.* Estimate Analysis of GARCH (2,1) Model

| Parameter | Estimate | Standard Error | t value | p-value |
|---|---|---|---|---|
| $c$ | $1.946\times 10^{-6}$ | $6.228\times 10^{-7}$ | 3.126 | 0.00177 |
| $\alpha_1$ | $1.970\times 10^{-1}$ | $3.983\times 10^{-2}$ | 4.946 | 0.0000* |
| $\alpha_2$ | $1.000\times 10^{-8}$ | $5.326\times 10^{-2}$ | 0.000 | 1.00000 |
| $\beta_1$ | $7.171\times 10^{-1}$ | $6.246\times 10^{-2}$ | 11.481 | 0.0000* |

*Actual p-value is lower than 0.0001

The estimate analysis for GARCH (2,1) model is shown in Table 8. Observing the results, not all of the p-value of the parameters shows significance at level 0.05. The p-value for $\alpha_2$ is greater than alpha, and indicates that GARCH (2,1) with parameters $c, \alpha_1, \alpha_2$, and $\beta_1$ is not a good fit for our data.

### 3.6 Model Selection

After fitting different models to the log returns ($Y_t$) of PSEi, we now use selection criteria to pick the final model among the models considered using Akaike's Information Criterion (AIC) and Log-likelihood (LL). Table 9 shows the values of the AIC and LL for the three models. The model that shows the minimum value of AIC and the maximum value for the LL will be selected as the best model for our data. GARCH (1, 2) model has the minimum value of AIC and has the maximum value of LL. This implies that GARCH (1,2) model is the most appropriate model among all the proposed models.

*Table 9.* Model Selection Criterion

| Models | AIC | LL |
|---|---|---|
| GARCH(1,0) | -10654.45 | 5329.224 |
| GARCH(1,1) | -10898.09 | 5452.046 |
| GARCH(1,2) | -10898.17 | 5453.087 |

### 3.7 Diagnostic Checking

To further check the adequacy of the selected model, the LM test is performed for the GARCH (1,2) model. Results of the test are presented in Table 10.

*Table 10.* LM Test for ARCH effect of GARCH (1,2) model

| Model | LM Test | |
|---|---|---|
| | $x^2$ | p-value |
| GARCH(1,2) | 15.108 | 0.2356 |

Observing the result, the p-value of the LM test for GARCH (1,2) residuals is greater than the level of significance, this implies that GARCH (1,2) fully captured the ARCH effect in the series and hence, the best model for our data.

Table 11. Kolmogorov-Smirnov Test for Normality

| **Kolmogorov- Smirnov Test** | | | | |
|---|---|---|---|---|
| **GARCH (1,2) Residuals** | **Mean** | **Standard Deviation** | **Value** | **p-value** |
| | -0.0485631 | 0.98897046 | 1.288 | 0.071 |

Kolmogorov-Smirnov Test is done to test the normality of the residuals of GARCH (1,2) model. Table 11 shows that the value of the test is 1.288 with the p-value equal to 0.071. Since the p-value is greater than 0.05 level of significance, this shows that the test fails to reject the null hypothesis and we conclude that the residual of the GARCH (1,2) is normal.

Table 12. Ljung-Box Test for Independence

| **Ljung-Box Test** | | | |
|---|---|---|---|
| **GARCH(1,2) Residuals** | $x^2$ | **df** | **p-value** |
| | 0.51046 | 1 | 0.4749 |

The Ljung Box test is done to check the independence of the residuals of GARCH (1,2) model. The result of this test is shown at Table 12. The Ljung Box test value of GARCH (1,2) is 0.51046 with the p-value 0.4749. Since the p-value is greater than the level of significance which is 0.05, this implies that the null hypothesis cannot be rejected and conclude that autocorrelation does not exist in the residual of GARCH (1,2).

## 4. DISCUSSION AND CONCLUSION

The daily Philippine Stock Exchange Composite Index shows an increasing growth from period 2011 to 2013. It can be observed that in the first months of 2013 there is a huge increase on the stock exchange, but in mid-2013 a sudden decrease was observed that continued for a few months. An increase and decrease of stock exchange is observed for a few months that continued until 2014. A sudden decrease in the exchange was also observed in mid-2015 but eventually never continued since an increase is observed afterward. The highest peak of the series can be seen in 2015 where the index reached to 4206.01. The PSEi ($P_t$) is non-stationary so we take its natural log and obtained a stationary log returns ($Y_t$). The ARIMA($p, d, q$) models were inappropriate to use for the log returns ($Y_t$) due to the presence of heteroskedasticity. GARCH ($r, s$) models was found to be an appropriate alternative since it can capture the ARCH effects. From all the possible GARCH models, the GARCH (1,2) is the best model since it has the lowest AIC value and has the highest LL value.